\documentclass[pre,onecolumn,preprint]{revtex4}

\usepackage[dvips]{graphicx}

\begin{document}

\preprint{}
\title{Anomalous diffusion in silo drainage }

\author{Roberto Ar\'{e}valo}
\affiliation{Departamento de F\'{\i}sica, Facultad de Ciencias, Universidad de Navarra,
E-31080 Pamplona, Spain.}
\author{Angel Garcimart\'in}\affiliation{Departamento de F\'{\i}sica, Facultad de Ciencias, Universidad de Navarra,
E-31080 Pamplona, Spain.}
\author{Diego Maza}
\affiliation{Departamento de F\'{\i}sica, Facultad de Ciencias, Universidad de Navarra,
E-31080 Pamplona, Spain.}

\pacs{{45.70.-n}{Granular systems} \and
{45.70.Mg}{Granular flow: mixing,segregation and
stratification}\and {45.50.-j}{Dynamics and kinematics of a
particle and a system of particles}}

\begin{abstract} 
The silo discharge process is studied by molecular
dynamics simulations. The development of the velocity profile and
the probability density function for the displacements in the
horizontal and vertical axis are obtained. The PDFs obtained at
the beginning of the discharge reveal non-Gaussian statistics and
superdiffusive behaviors. When the stationary flow is developed,
the PDFs at shorter temporal scales are non-Gaussian too. For big
orifices a well defined transition between ballistic and diffusive
regime is observed. In the case of a small outlet orifice, no well
defined transition is observed. We use a nonlinear diffusion
equation introduced in the framework of non-extensive
thermodynamics in order to describe the movements of the grains.
The solution of this equation  gives a well defined relationship ($\gamma = 2/(3-q)$) 
between the anomalous diffusion exponent $\gamma$ and 
the entropic parameter $q$ introduced by the non-extensive formalism to fit the PDF of the
fluctuations. 
\end{abstract} 

\volumeyear{}
\volumenumber{}
\issuenumber{}
\eid{}
\date{24 May 2007}
\startpage{1}
\endpage{}
\maketitle

\section{Introduction}
\label{intro}

The process of gravity-driven silo discharge could naively be
expected to be a simple one, yet we lack a well defined theoretical
framework to explain the experimentally observed grain dynamics in
terms of fundamental interactions. During the silo drainage,
particles evolve under the action of gravity and interact between
them through inelastic collisions in a complicated way, where
intensive variables --such as density or temperature-- are not well
defined. Under these circumstances, the displacement and velocity
fluctuations could be of the same order than its mean values, and
the net force acting on each particle could even present strong
deviations compared to the gravity force, therefore introducing
unwanted effects like segregation or jamming
\cite{segregation,To,Nosotros2}. Two models were generally
considered successful in explaining the global characteristics of
the flow (the velocity profile) inside the silo. One of them is
based on a continuous approach and the other takes into account the
discrete nature of the media.  The continuum model \cite{Nedderman1}
uses concepts like elasto-plastic potentials introduced in the
framework of continuum mechanics, in which the mean velocity field
can be obtained. Nevertheless, this approach looses validity near
the outlet orifice and does not bring any information about possible
\emph{microscopic} effects like mixing or segregation. The second
one is based on the idea that each individual particle executes
haphazard movements which can be treated as a random walk. This
notion was originally introduced by Litwiniszyn in the sixties and
is commonly called \emph{diffusive void model} \cite{Litw1}. This
model focuses on the movement of voids injected at the outlet of the
silo. Such voids diffuse upwards, exchanging their positions with
the grains, which in turn move towards the orifice at the bottom. In
spite of their different premises, both approaches give essentially
the same results for the mean velocity profile, but in the latter
case the profile depends only on a single parameter, that can be
regarded as a characteristic length $\alpha$ related to diffusion.

In this work we present numerical simulations of a silo discharge
process, including the beginning of the operation. We use three
outlet diameters to study the behavior of the velocity profile,
demonstrating an evolution between the transitory and stationary
states. Results for the PDFs of the displacements of individual
grains reveal non-Gaussian statistics and super-diffusive behavior
at the beginning of the discharge. In agreement with experiments,
non-Gaussian to Gaussian PDF's transition is observed in the
stationary regime. Finally, we show that the complete sequence of
dynamical states displayed by the particles at the beginning of the
discharge can be interpreted in the non-extensive statistical
mechanics framework introduced by Tsallis \cite{Tsallis1}, which can
be used to fit the obtained PDFs if the anomalous scaling of
mean-square displacement as a function of time is taken into
account.

\section{The diffusive approach}
\label {diffusive}

The diffusive void model yields a well defined relationship between
the fluctuations of the positions of the particles and the
characteristic length used to fit the velocity profile
\cite{Litw1,Mullins}. But recent experiments have evidenced some
discrepancies concerning these predictions. In \cite{Choi1}, a
high-resolution particle-tracking experiment of a silo discharge is
reported. The estimated value for the parameter $\alpha$ --around
$2$ or $3$ particle diameters-- is much larger than the one
predicted by the aforementioned authors. Choi \emph{et al.} also
reported the observation of non-Gaussian statistics and a
super-diffusive regime of the grain displacements at sufficiently
short temporal scales. (This temporal scale is defined by the time
it takes for a bead to fall a distance of its own diameter). The
probability density functions (PDFs) obtained for the particle
displacement in the vertical and horizontal directions at this scale
are both fat-tailed. For larger displacements both PDFs evolve
toward a Gaussian shape. The grains undergo a transition from a
super-diffusive to a diffusive regime in a coarse grained scale. At
these scales, the velocity profile is stable and the diffusive
models seem to regain their validity. More recently, a similar
experiment was reported  where the non-Gaussian fluctuations remain
even for the coarse grained scale \cite{moka}.

\begin{figure}
\includegraphics{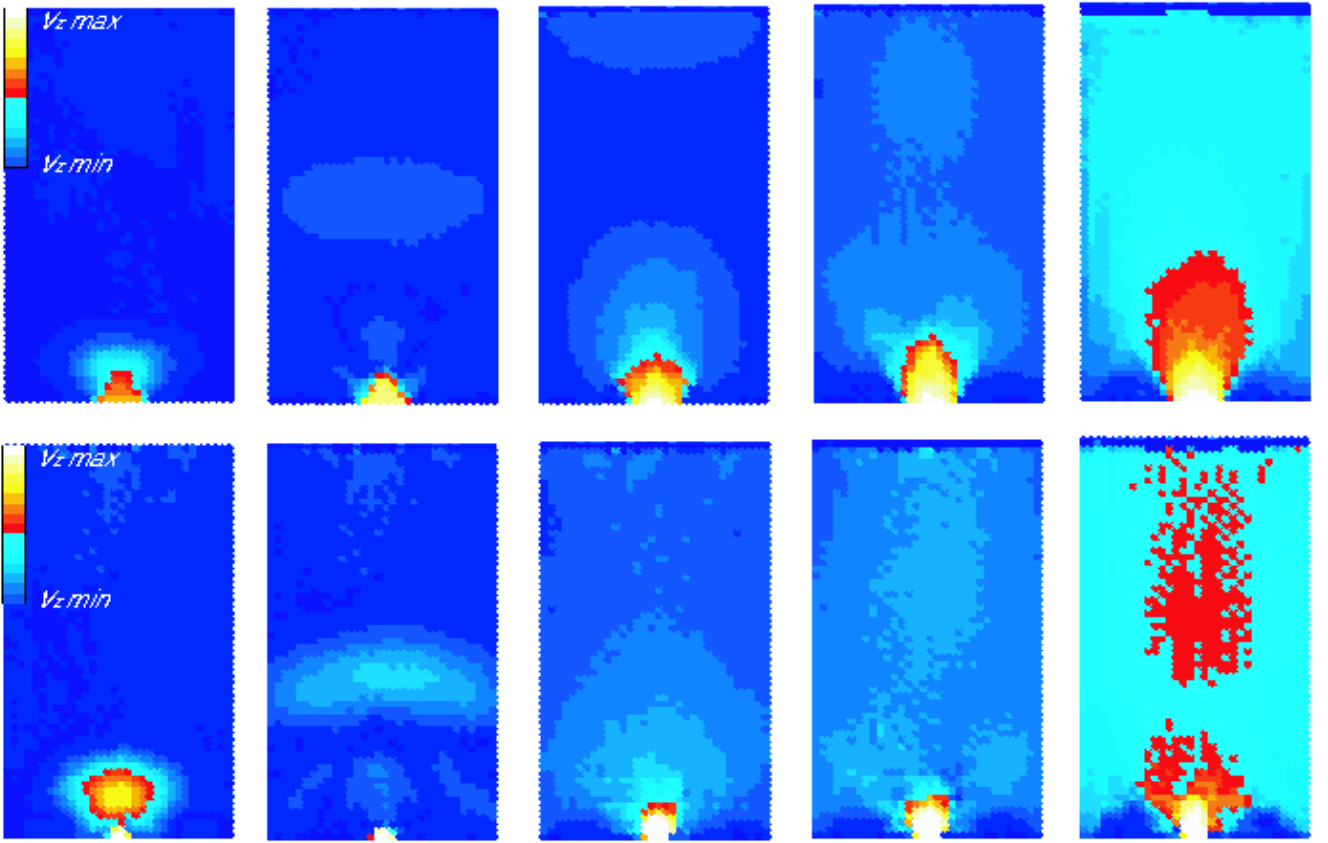}
\caption{Average vertical velocity profile inside the silo for
orifices of size $11$   (\emph{top}) and $3.8\,d$ (\emph{bottom}) at
different stages of its evolution. Time increases from left to
right. The last picture corresponds to the fully developed regime.
The localized structures correspond to a well defined region where
strong velocity gradient exist. Note than this zones does not
totally  disappear in the stationary regime of the small outlet
orifice. } \label{fig1}
\end{figure}

Let us to go through the fundamental assumptions of the diffusive
models. The line of reasoning proposed  by Litwiniszyn \cite{Litw1},
later developed by Mullins \cite{Mullins}, basically introduces two
fundamental assumptions: (i) the particle moves following a random
walk in which the interaction between horizontal an vertical
displacements can be neglected, and (ii) the time elapsed in one
jump is almost the same than the one corresponding to a free fall.
Assuming these hypothesis, in a typical experimental situation the
diffusivity of the particle scales with its diameter $d$ as
$D\simeq\sqrt{gd^{3}}$, and the parameter $\alpha$ --which Mullis
named ``jumping distance''-- can be expressed as $D/V=3/8d$, where
$V$ is the mean velocity in the vertical direction. The parameter
$\alpha$ is relevant to check the theory against the experimental
results because it is all that is needed to fit the velocity
profile. The same parameter was derived from probabilistic arguments
for an hexagonal array of particles \cite{Litw1}, yielding the value
$d/\sqrt{6}$. Taking into account the strongly biased character of
the diffusive movements in the vertical direction, it is possible to
write an analytic expression for the velocity profile (assuming a
point source of voids) \cite{Mullins}:

\begin{equation}
v(r,z) \propto \,exp\,\left(-\frac{r^{2}}{4 \alpha z}\right)
\label{sol_mullins}
\end{equation}

where $r^{2}=x^{2}+y^{2}$.

The same profile was derived from phenomenological arguments by
Nedderman $\&$ T\"uz\"un at the end of the the seventies
\cite{Tuzun}.  Following a kinematic reasoning, this model
reproduces quite well the velocity profile inside the silo by
assuming that the horizontal component of the velocity $u$ is
straightforwardly proportional to the gradient of the vertical
velocity $v$:

\begin{equation}
u=B\frac{\partial v}{\partial x} \label{lev}
\end{equation}

Considering the granular media as an incompressible material, an
equation can be deduced for the velocity profile which is isomorphic
to the one obtained following the diffusive model:

\begin{equation}
\frac{\partial v}{\partial z}=B\frac{\partial ^2v} {\partial x^{2}}
\label{lde1}
\end{equation}

In this \emph{kinematic model}, the temporal variable has been
replaced by the coordinate $z$. The solution to this parabolic
equation corresponds to a Gaussian profile whose shape only
depends on the parameter $B$:

\begin{equation}
v(x,z)=\frac{Q} {\sqrt{4 \pi B z}} e^{-x^{2}/4Bz} 
\label{lde2}
\end{equation}

where $Q$ stands for the volumetric flow.

The parameter $B$ appearing in Eq.~\ref{lde1} plays the same role
than the ``jumping length'' $\alpha$ introduced by diffusive models.
But contrary to them, the kinematic model does not provide any
explanation about the dependence of $B$ on the typical variables of
the problem --such as particle diameter, silo size or any other
characteristic length. In all the cases reported, the value of $B$
lies between $2$ or $3$ times the particle diameter. Besides, $B$
depends on the position inside the silo \cite{Choi2}.

Although the experimentally determined value of $B$ does not match
the jumping length predicted by the diffusive models either, both
approaches rely on just a single parameter to describe the flow.
This makes these kind of models appealing when trying to describe
the movements of the particles in the discharge process.

Recently, a more general description was introduced \cite{Bazant1}
in order to provide an explanation for the discrepancy between the
experimentally determined figures for the ``diffusive length'' and
the values predicted by the diffusive approach. This new model
assumes that the voids injected at the outlet of the silo do not
match the size of just one particle. Instead, the injected void
spreads through several grains, which will then move cooperatively.
The resulting ``cluster'', which spreads the void throughout a group
of particles, is called a ``spot''.  Spots move upward inside the
silo and the grains affected by the spot carry out small movements
toward the base. As the grains affected by the spot must move in a
concerted fashion, it is not surprising that the displacements of
nearby particles present strong correlations. The resulting
\emph{spot diffusivity} allows to recover the values for the jumping
length obtained in experiments.

Let us note that in all the cited references the discharge process
must have reached a steady state regarding the flow, far enough from
the beginning of the discharge. Moreover, the size of the orifice
was large enough to avoid jamming events like those described in
\cite{Nosotros}. Remarkably, the universality of the results may be
lost when the size of the outlet orifice is small (see the case of
small flow in reference \cite{Choi1}). Regardless of the inherent
complexity of these states, a diffusive approach might be suitable
to describe the particle displacements, although the simple
assumptions introduced by the model could lose their validity. Early
stages of the discharge process involve a densely packed media
subjected to a strong, inhomogeneous shearing. Furthermore, even in
the case of homogenous shear, densely packed granular media is prone
to complex dynamics, as reported in \cite{Kurchan,Kurchan2,Radjai},
where concepts like particle mobility and diffusivity to are invoked
to describe the particle dynamics.

\section{Numerical simulations}
\label{sec:Num_Sim}

We have carried out molecular dynamics simulations of disks in two
dimensions \cite{Schaffer1}. A simulation begins with $5000$ disks
arranged in a regular lattice; they are then given random
velocities, which have a Gaussian distribution. The disks are
allowed to fall under gravity through a conical silo. Below this
hopper lies a flat bottomed silo where the grains are deposited.
This is the preparation phase, which is purposely long and complex
in order to break the correlations that the initial regular
arrangement of the grains might induce in the dynamics. Once all the
grains have fallen into the flat silo we wait until most of the
kinetic energy is dissipated. Finally, we open an outlet at the
bottom allowing the grains to fall and we start our measurements.

The walls of the two silos used are built with grains. The
interaction between grains is the same than the interaction between
grains and walls, but the latter are fixed in their places. Thus the
walls are rough and cause dissipative collisions. The width of the
silo base is $50$ grain diameters and the height reached by the
grains when the silo is full is approximately twice that length.
These dimensions guarantee that the flow rate does not depend on
wall or filling effects. In addition, the filling height changes
only slightly during a simulation.

The model for the forces describing the interaction of two particles
$i,j$ consists of normal and transversal contacts as well as
dissipative terms:

\begin{eqnarray}
F_{n}=k_{n}\xi^{3/2}-\gamma_{n}v_{i,j}^{n} \label{normal}\\
F_{t}=-\min\left(\mu\vert F_{n}\vert, \gamma_{s}\vert
v_{i,j}^{s}\vert\right) \cdot \textrm{sign} \left(v_{i,j}^{s}\right)
\label{tangent}
\end{eqnarray}

where
\begin{equation}
v_{i,j}^{s}=\dot{\textbf{r}}_{ij}\cdot \textbf{s}+\frac{1}{2}
d\left(\omega_{i}+\omega_{j}\right) \label{shvel}
\end{equation}

Eq.~(\ref{normal}) gives the force in the normal direction of the
contact. The first term is a restoring force proportional to the
superposition $\xi$ of the disks. The $3/2$ exponent arises from the
Hertz theory of the contact. The second term is a dissipation
proportional to the relative normal velocity of the interacting
disks with damping coefficient $\gamma_{n}$. Eq.~(\ref{tangent}) is
the sliding component of the damping force. It implements the
Coulomb criterion with friction coefficient $\mu$. The damping in
the transverse direction is proportional to the shear velocity given
by Eq.~(\ref{shvel}) and a transverse damping coefficient
$\gamma_{s}$. In this equation $\textbf{s}$ is a unit vector
tangential to the disks at the contact point, and $\omega_{i}$ and
$\omega_{j}$ are their angular velocities. Thus, in this scheme of
forces we have a restoring force which prevents grains to
interpenetrate (although a very small penetration is needed) along
with damping terms in the normal and tangential directions which
dissipate energy during the contact. This dissipation is among the
most prominent characteristics of granular media.

The values of the coefficients are, in reduced units,
$k_{n}=10^{5}\, mg/d$, $\gamma_{n}=100\, m\sqrt{g/d}$,
$\gamma_{s}=300\,m\sqrt{g/d}$, and $\mu=0.5$. The integration
time-step used is $1.25^{-4}\,\tau$ with $\tau=\sqrt{d/g}$, and $m$,
$d$ and $g$ stand respectively for the mass and diameter of the
disks and the acceleration of gravity.

The equations of motion were integrated using the velocity-Verlet
algorithm and we used a neighbor list \cite{Rapaport} to reduce the
computational effort.

\subsection{Velocity profiles and probability density functions}
\label{sec:2}

Using three different diameters of the exit orifice (namely,
$3.8\,d$, $11\,d$ and $16\,d$) we studied the evolution of the
vertical velocity profile. The smaller and larger diameters where
chosen because they belong clearly in two distinct regimes: for the
former, the flow can be intermittent, while for the latter it is not
\cite{Nosotros}. We compute the velocity profile from the moment
when the outlet is opened until the moment when the velocity profile
becomes stationary. For each orifice size we perform $20$
independent simulations and average them to obtain the final result.

\begin{figure}
\includegraphics[scale=0.8]{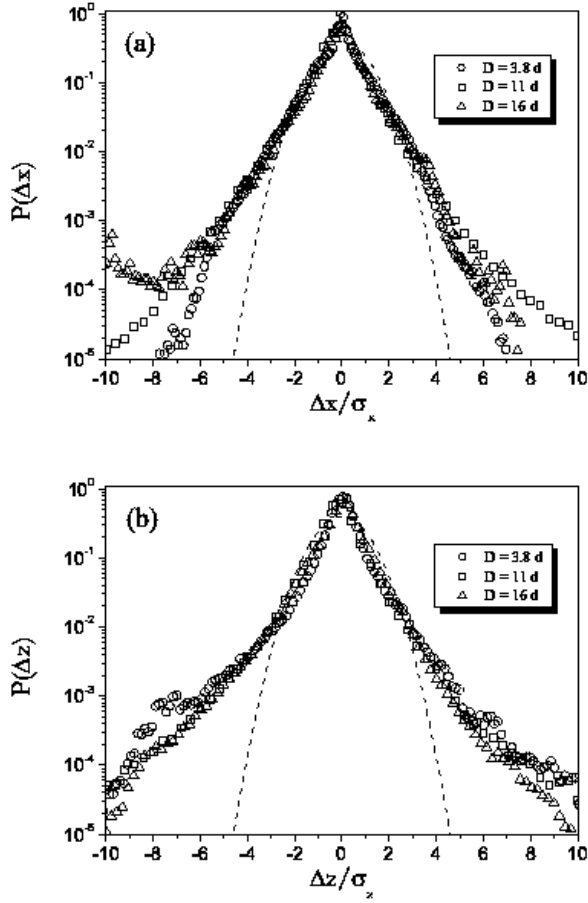}
\caption{ Normalized PDFs for (a) horizontal and (b) vertical
displacements. These distributions have been obtained in a temporal
window where the averaged particle displacement is \emph{lower} than
a particle diameter. The dotted line is a Gaussian.} \label{fig2}
\end{figure}

\begin{figure}
\includegraphics[scale=0.8]{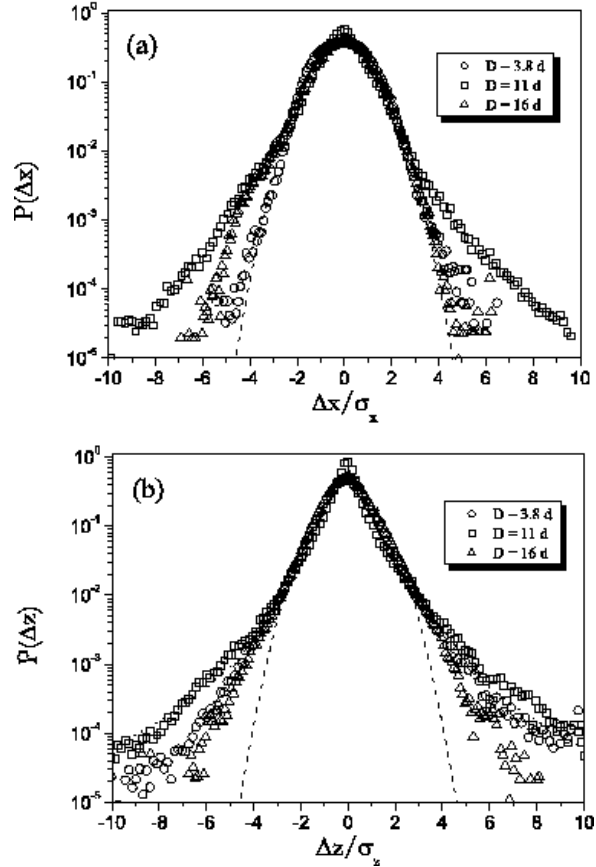}
\caption{Normalized PDFs for (a) horizontal and (b) vertical
displacements. These distributions have been obtained in a temporal
window where the averaged particle displacement is \emph{larger}
than a particle diameter. The dotted line is a Gaussian. }
\label{fig3}
\end{figure}

In fig.~\ref{fig1} we show the evolution of the averaged velocity
profile (only its vertical component) for the $3.8$ and $11\,d$
orifices. It is evident that in both cases groups of grains move
downward together at the very beginning of the discharge, while
structures that can be described as ``bubbles'' can be seen moving
upward. Let us stress that these spatial structures do not
correspond exactly to the spots introduced in \cite{Bazant1}. These
bubbles are zones where the mean velocity is larger than the bulk
and its evolution reveals the intermittent regime at the beginning
of the discharge. When these bubbles disappear, the characteristic
stable flow profile is developed and the asymptotic velocity
converges to a Gaussian profile. In the asymptotic regime (for the
bigger orifices) the velocity profile in the middle of the silo can
be fitted using a numerical ``diffusivity length'' of $2.2\,\pm\,
0.2\,d $ which is in excellent agreement with experiments
\cite{Choi1,Tuzun}.

The results obtained with the smaller exit orifice are similar but
an important difference arises: the characteristic time needed to
reach a stable flow grows dramatically as the outlet diameter
decreases. This feature is related to the fact that an increasing
number of bubbles appear in the system, inducing an intermittent
flow in the silo. Under these circumstances, strong velocity
gradients can be observed near the exit orifice. This implies that
some assembly of particles can have a relatively low velocity that
may enable them to form an arch near the outlet, and stable jamming
events can arise if the particles arrest the flow. This intermittent
regime will be studied anywhere.

\begin{figure*}
\includegraphics[width=1.0\textwidth]{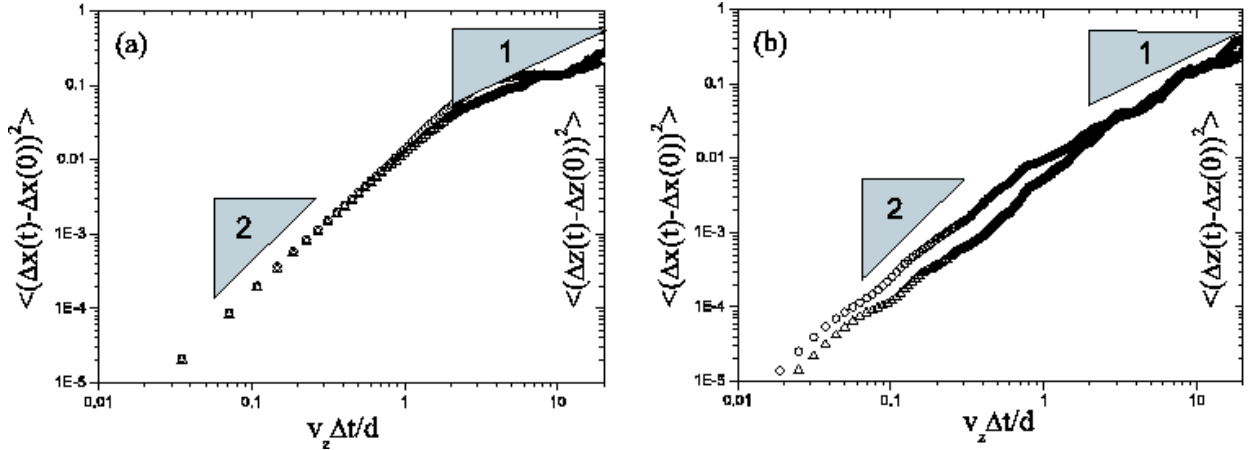}

\caption{ Mean squared displacements for the horizontal and vertical
displacements in logarithmic scale. Circles correspond to the $x$
component and triangles to the $z$ component. (a) $16\,d$ silo. The
ballistic character for small displacements is obvious; once the
displacement exceeds a distance of about the diameter of the
particle the movement becomes diffusive. (b) $3.8\,d$ silo. For
small exit orifices it is difficult to distinguish the two behaviors
and the crossover at about one $d$ mentioned for the large orifice.
The movement of the particles seems to be superdiffusive at all
scales.}

\label{fig4}
\end{figure*}

We measured the displacements of individual grains during the
discharge. In order to compare our results with those obtained
experimentally, we chose a sampling time such that the mean
displacement of the particles during it is of about $0.01\,d$ (which
is approximately the same than in the experiments). The particles
were tracked in a window with different dimensions ($10$x$10$,
$15$x$15$ and $20$x$20$ $d$) at the center of the silo. We obtain
the same results regardless of window size. When computing the PDFs
of the vertical displacements we subtract the corresponding
component of the mean flow velocity.

In fig.~\ref{fig2} we show the PDFs corresponding to the fully
developed flow in semilogarithmic scale. The displacements in each
direction are normalized by their standard deviation. We see that
the PDFs are essentially the same for the vertical and horizontal
displacements. Besides, they are clearly non-Gaussian, with apparent
differences both in the central region and in the tails, which are
fat. As reported in \cite{Choi1}, the fluctuations in the horizontal
direction evolve toward a Gaussian profile when the displacements
are larger than the diameter of the particle. On the other hand, the
PDF corresponding to the fluctuations in the vertical direction
remains long-tailed even for longer periods of time (see
fig.~\ref{fig3}.b) wich is consistent with the experimental result
reported by Moka \emph{et. at.} \cite{moka}.

In fig.~\ref{fig4} we show the mean squared displacements in each
direction as a function of particle displacement (normalized by the
particle diameter $d$) for a small and a large orifice. In the case
of the largest orifice (fig.~\ref{fig4}.a), a well defined crossover
between two different regimes is displayed. For displacements up to
about one particle diameter, the variance displays a ballistic
behavior, as in molecular fluid transport \cite{liquidos}. A normal
diffusive regime can be used to describe particle fluctuations equal
or larger than one diameter for big orifices.

On the contrary, for the small orifice (fig.~\ref{fig4}.b) there is
not a well defined transition from ballistic to diffusive regime.
Furthermore, subparticle displacements are superdiffusive. This is
consistent with the fact that the corresponding PDF (fig.
\ref{fig3}) remains fat tailed at all scales. Remarkably, a
subparticle superdiffusive regime was reported in the experiment
\cite{Choi1} for all the outlet orifices studied. Such an effect
would be associated to the strong influence of the lateral wall on
the particle dynamics where the particle tracking is performed.

It is remarkable that a single parameter like the one introduced by
the kinematic model \cite{Tuzun} or the jumping length introduced by
Mullins \cite{Mullins} can be used to obtain the shape of the
velocity profile, at least for large enough orifices. Unfortunately,
the diffusive length $B$ introduced by simple phenomenological
arguments can not be easily associated with the microscopic grain
dynamics. The diffusive model also predicts a Gaussian velocity
profile that depends on a single parameter. Seemingly the meaning of
this parameter is the same: a ``characteristic diffusive length'';
but care is needed in order to compare both two. The parameter
introduced by Mullins is the ratio between diffusivity and the
characteristic velocity of the grains (or the voids, for that case)
during a ``jump'' comparable to its own diameter, whereas $B$ is
just a phenomenological parameter introduced to link the two
velocity components (as no prediction is provided by the model, it
must be obtained from experiments). But there is not any \textit{a
priori} a reason to consider both parameters equivalent from a
theoretical point of view.

Admittedly, the predictions of the diffusive models (including the
void model, which assumes the unrealistic situation of the voids
executing a simple Bernoulli random walk in a regular lattice) do
not match quantitatively the fitting parameter for the velocity
profile found in numerical simulations or experiments. As noted
above, a new model was introduced to tackle this discrepancy
\cite{Bazant1}. This model proposes a cooperative mechanism or
``cage effect'' to represent the particle void interaction. This
effect gives a correct estimation of the velocity fitting parameter
but does not explain the microscopic diffusive regime.

The spot model infringes one of the most important hypothesis of the
diffusive models: void and particle displacement are no longer
symmetrical. The symmetry plays a crucial role in the calculation of
particle diffusivity by determining the time that a grain needs to
migrate a distance equal to its own diameter. When a void enters the
silo, the particle must perform a ballistic flight in a time scale
comparable to the free fall time. The numerical simulation confirms
the ballistic flight in the fully developed flow (Fig~\ref{fig4})
but in a temporal scale considerable larger than a free fall. In our
simulations the flight time is three or four times longer than the
free fall time, depending on the position in the silo. This spatial
dependency also agrees with experimental results \cite{Choi2} that
reveal a variation of the parameter $B$ with the particular place of
the silo where it is measured. Working is in progress in this line
and results will be discussed elsewhere.

\section{A non-linear diffusive approach for the beginning of the discharge}

\begin{figure*}
\includegraphics[width=1.0\textwidth]{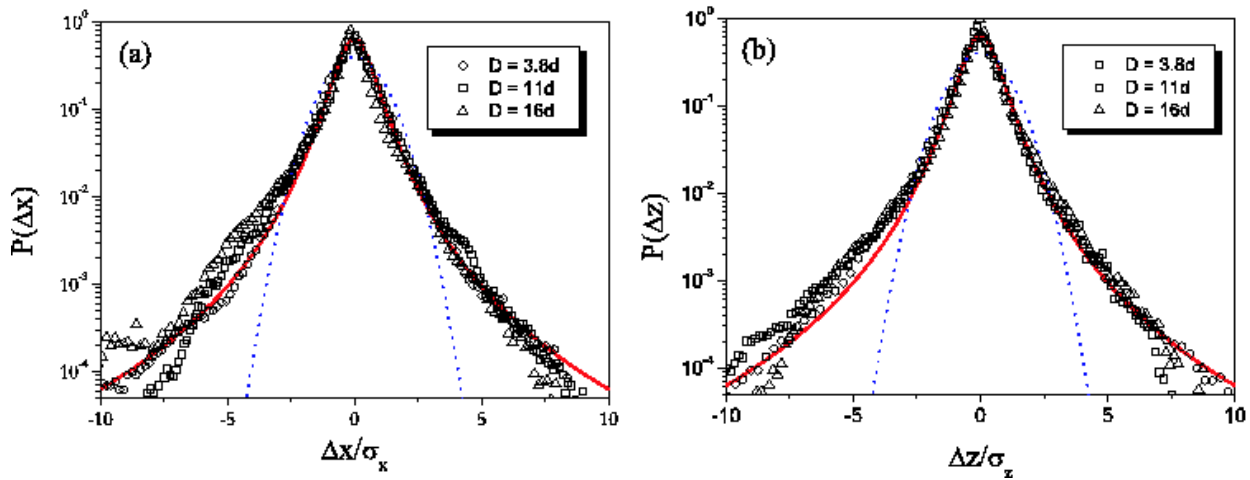}

\caption{ Normalized PDFs for (a) horizontal and (b) vertical
displacements at the early stages of the discharge. These
distributions have been obtained in a temporal window spanning
from the beginning of the discharge to the moment when the mean
displacement is twice the diameter of the beads. The dotted line
is a Gaussian. The continuous line is Eq.~\ref{pdf}  }
\label{fig5}
\end{figure*}

Let us now describe the particle dynamics at the beginning of the
discharge. This is important in order to understand the origin of
jamming and the existence of a critical radius \cite{Nosotros}
beyond which the flow is never interrupted.  In fig~\ref{fig5} the
normalized PDFs for the displacements in each direction are
displayed. They are clearly non-Gaussian and differ a little bit
from the ones obtained for a stationary flow. This is indeed as
expected. At the early stages of the discharge, the particle motions
should be more correlated than afterwards. The material needs to
dilate in order to flow, and the dilation takes place preferably
along the regions where more empty volume is available. Under these
circumstances, contact network or force chains can persist for a
finite time and eventually block the outlet orifice. At the
beginning of the discharge, the packing fraction will be lower than
the corresponding to the ideal close packing of an hexagonal lattice
and an important number of arches \cite{Luis} will introduce long
range interactions between particles. At the same time, each
particle will be affected by non trivial stresses along the vertical
and horizontal directions.

\begin{figure*}
\includegraphics[width=1.0\textwidth]{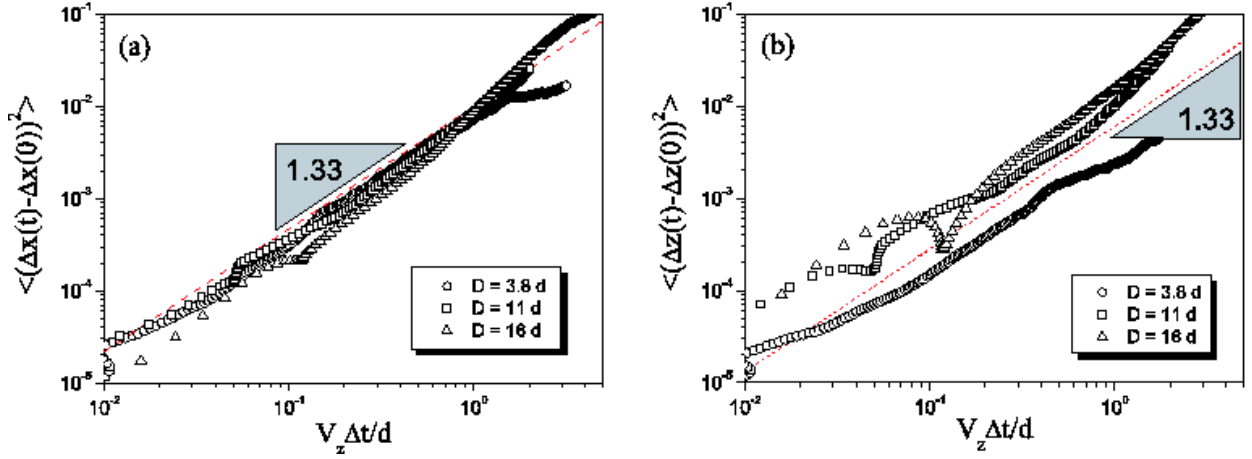}
\caption{  Mean squared displacements for the horizontal (a) and
vertical (b)  displacements at early stages of the discharge. The horizontal
component shows clearly a superdiffusive behavior in all the cases.
The vertical component displays a similar behavior although it is
not as obvious. The curves plotted would become smoother if the
number of averaged discharges is increased.}
\label{fig6}
\end{figure*}

The mean squared displacements  scale approximately as $\left<\Delta
x^{2}\right> \cong \left<\Delta z^{2}\right>\propto t^{4/3}$ for all
the orifices studied (Fig.~\ref{fig6}). Moreover, there is not a
clear crossover between superdiffusive to diffusive regime even for
big orifices. These results agree with the experimental observation
that even for large orifices is necessary to wait an easily
mensurable time before reaching the stationary regime. Clearly this
time strongly depends on the orifice size.

The anomalous scaling and the PDFs obtained at the beginning of the
discharge seem to be universal and suggest that it is necessary to
introduce a generalized expression to describe the ``walk'' or
displacements of the particles.

Some theoretical models have been introduced to describe these
familiar features of granular media using alternative non standard
thermodynamical descriptions, which resort to concepts like granular
temperature \cite{Edwards}. But these dynamical states are in fact
quasi-stationary \cite{Kurchan,Kurchan2} whereas in our case it
would not be pertinent to apply a concept like temperature at the
beginning of the discharge. Besides, recent numerical \cite{Radjai}
and experimental works \cite{Bob} have shown that when a system of
grains is affected by shearing effects, anomalous scaling
relationships are general. As argued by many authors, any transport
mechanism where the variance of particle displacements scales with
time as $t^{\gamma}$ with $\gamma\neq1$ is connected to anomalous
diffusion \cite{Fokker,zanete}. A great variety of physical systems
can display fluctuating transport mechanisms where such anomalous
scaling can be found \cite{libro}. The paradigmatic example is a
random walk performed in a disordered network \cite{Fokker}. In such
a case, randomness in the lattice induces inhomogeneous transition
rates and as consequence the diffusive process is anomalous. In our
simulations, the PDFs for the displacements indicate that the
particle diffusion at the beginning of the discharge is anomalous.
We will now introduce a formal framework to describe the anomalous
behaviors.

There are two important attempts to generalize the normal diffusive
equation to include anomalous dynamics. The first one remains linear
but introduces the concept of \emph{fractional derivative} in its
formulation \cite{Compte}. This formalism will be not discussed here 
due to the fact that it does not display a transition to normal diffusion 
at longer times. The second one is to switch to a \emph{nonlinear} equation (NLDE)
\cite{Plastino}; in the simplest version it reads:

\begin{equation}
\frac{\partial p\left(x,t\right)}{\partial t}=D\frac{\partial
^2\left[p\left(x,t\right)\right]^\nu} {\partial x^{2}} \label{nlde}
\end{equation}

where $\nu$ is a real number. One of the consequences of this
functional dependence is that the diffusivity has now a complex
dependence on the density probability function $p(x,t)$.  The
microscopic origin of such an equation has been discussed by many
authors \cite{Plastino,Jou,Borland}, specially in the framework of
the Tsallis entropy formalism \cite{zanete,libro}. Such formalism
was proposed to analyze long range interacting systems and its
non-extensive properties. Since its introduction,  Tsallis
statistics has been used in a broad range of physical problems where
non-Gaussian PDFs and anomalous diffusive behaviors appeared
\cite{Upadhyaya,Bodenschatz,Renzoni}, even in the framework of a
granular gas \cite{Sattin}.

Tsallis formalism is important here because it gives an analytic
solution to Eq.~(\ref{nlde}) introducing a new parameter $q$ called
\emph{entropic parameter} \cite{zanete}. Considering a 1D, case the
solution for this equation is given by:

\begin{equation}
p(x,t)=\frac{A_{q}}{\sqrt{3-q}(Dt)^{1/(3-q)}}e_{q}^{-x^{2}/[(3-q)(Dt)^{2/(3-q)}]}
\label{nlde2}
\end{equation}

where $q=2-\nu<3$. It has been shown \cite{zanete} that the scaling
between time and mean square distance can be expressed as a function
of $q$ through the expression $\left<x^2\right>\propto t^{2/(3-q)}$,
provided that the second moments of the displacement distribution
remain finite, which is our case.

Let us consider the case where $\left<x_{2}\right> \cong
\left<z^{2}\right>\propto t^{4/3}$ (Fig. ~\ref{fig6}). Comparing
this numerically obtained relationship and the former equation, we
can estimate the value for the $q$ parameter: $q=3/2$. With this
value, we can write the analytical expression for the solution of
Eq.~(\ref{nlde}) normalized by its standard deviation \cite{libro}
as:

\begin{equation}
p\left(x\right)=\frac{2/\pi}{\left(1+x^{2}\right)^{2}} \label{pdf}
\end{equation}

This function corresponds to the solid line plotted in
fig.~\ref{fig2}. The agreement between the numerical result and the
analytical prediction indicates that a NLDE like Eq.~\ref{nlde}
represents fairly well the evolution of the particles' fluctuations
both in the vertical an horizontal directions at the early stages of
the discharge.

Furthermore, as time increases the solutions provided by the Tsallis
formalism to Eq.~(\ref{nlde}) tend to a Gaussian for values of $q <
5/3$ \cite{libro}, as observed in our simulations when the system
reaches the stationary regime (for the $x$ component). Although the
meaning of the diffusivity constant in normal diffusion can be
related to well defined concepts like temperature and mobility, the
diffusivity introduced in Eq.~(\ref{nlde}) lacks any obvious
physical meaning. At this stage of the discharge, the diffusivity
$D$ would be determined by the complex interactions between flowing
particles. They give rise to a transport coefficient with a physical
meaning unavoidably more complex than just a characteristic length
obtained from a well defined particle velocity.

\section{Discussion}

We have studied the temporal fluctuations of particle displacements
in the discharge of a silo by gravity. We show how at the early
stages of the discharge the particle displacement presents
distinctive features, as previously suggested by many authors. For
that regime, it is possible to derive the PDFs for the displacement
fluctuations of the particles from the non-extensive entropy theory
introduced by Tsallis. We have shown that this formalism allows the
treatment of diffusive systems presenting anomalous behavior through
the introduction of a nonlinear diffusive equation. This equation
would then provide a formal framework to represent the dynamical
evolution of the PDFs from the beginning of the discharge to the
fully developed regime. The introduction of this formalism might
also offer an explanation for the jamming probability introduced in
previous works \cite{Nosotros,To}.

For the stationary regime, we corroborate the non-Gaussian features
of the particle fluctuations for small displacements. Nevertheless,
we found that the subparticle diffusive motions are ballistic, as is
usual for the molecular fluids. In our opinion, the superdiffusive
regime reported in some experiments \cite{Choi1} could be related to
the effects of lateral walls on the particle mobility. It is a
widely accepted fact that the typical boundary layer in granular
material spans around $10$ particle diameters, so the lateral size
of the silo could cause strong effects on the particle displacement.

The transition between ballistic to normal diffusive regime
validates the applicability of diffusive models. The discrepancy
observed between the characteristic length used to fit the velocity
profile and the one predicted by the model is caused by a wrong
estimation on the scale for the typical subparticle displacement
velocity. The non-Gaussian nature for the PDFs for the vertical
coordinate should be certainly linked with this scales and will be
studied in the future.

\section{Acknowledgments}

This work has been supported by project FIS2005-03881 (MEC,
Spain), and PIUNA (University of Navarra). R.A. thanks Friends of
the University de Navarra for a scholarship.


\begin{thebibliography}{99}

\bibitem{segregation}
   A. Samadani, A. Pradhan, \& A. Kudrolli, Phys. Rev E \textbf{60}, 7203 (1999).

\bibitem{To}
   K. To, P-Y Lai, \& H. K. Pak, Phys. Rev. Lett. \textbf{86}, 71 (2001).

\bibitem{Nosotros2}
    I. Zuriguel, L. A. Pugnaloni,  A. Garcimart\'in \& D. Maza,
     Phys. Rev. E \textbf{68}, 030301(R) (2003).

\bibitem{Nedderman1}
   R. M. Nedderman, \textit{Statics and Kinematics of granular
   Materials},
   {Cambridge Univ. Press, Cambridge UK, 1992.}

\bibitem{Litw1}
   J. Litwiniszyn, Bull Acad. Pol. Sci. \textbf{11}, 593 (1963).

\bibitem{Tsallis1}
   C. Tsallis,
   J. Stat. Phys. \textbf{52}, 4579 (1998).

\bibitem{Mullins}
   W.W. Mullins,
   J. Appl. Phys.,\textbf{43}, 665 (1972).

\bibitem{Choi1}
   J. Choi, A. Kudrolli, R. R. Rosales, \and M. Z. Bazant,
   Phys. Rev. Lett. \textbf{92}, 174301 (2004).

\bibitem{moka}
   S. Moka, P. R. Nott, Phys. Rev. Lett. \textbf{95}, 068003 (2005).

\bibitem{Tuzun}
   R. M. Nedderman, U. T\"uz\"un,
   Powder Technology \textbf{22}, 243 (1979).

\bibitem{Choi2}
   J. Choi, A. Kudrolli \and M. Z. Bazant, J. Phys-Cond. Matt.
\textbf{17}, S2533 (2005).

\bibitem{Bazant1}
   M.Z. Bazant,
   Mechanics of Materials \textbf{38}, 717 (2006).

\bibitem{Nosotros}
   I. Zuriguel, A. Garcimart\'in, D. Maza, L. A. Pugnaloni \&  J. M.
   Pastor, Phys. Rev. E \textbf{71}, 051303 (2005).

\bibitem{Kurchan}
   H. Makse  \& J. Kurchan,
   Nature \textbf{145}, 614 (2002).

\bibitem{Kurchan2}
   B. Utter \& R. P. Behringer,
   Phys. Rev. E \textbf{69}, 031308 (2004).

\bibitem{Radjai}
   F. Radjai \& S. Roux,
   Phys. Rev. Lett. \textbf{92}, 174301 (2004).

\bibitem{Schaffer1}
   J. Sch\"afer, S. Dippel \&  D. E. Wolf,
   J. Phys. I(France) \textbf{6}, 5 (1996).

\bibitem{Rapaport}
   D.C. Rapaport, \textit{The art of molecular dynamics simulation},
   {Cambridge University Press, Cambridge UK 2004}.

\bibitem{liquidos}
   C. S. Campbell,
   J. Fluid Mech. \textbf{348}, 85 (1997).

\bibitem{Luis}
   R. Ar\'evalo , D. Maza  \& L. Pugnaloni,
   Phys. Rev. E \textbf{74}, 0213031 (2006).

\bibitem{Edwards}
 S. F. Edwards,
 \textit{Granular Matter: An Interdisciplinary Approach},
 121, {Springer, New York, 1994}.

\bibitem{Bob}
  J. Geng \& R. P. Behringer, Phys. Rev. Lett. \textbf{93}, 238002 (2004).

\bibitem{Fokker}
   J.-P. Bouchaud  \& A. Georges,
   Physics Reports \textbf{195}, 127 (1990).

\bibitem{zanete}
   D. H. Zanette,
   Brazilian Journal of Physics \textbf{29}, 108 (1999).

\bibitem{libro}
 \textit{L\'evy Fligths and Related Topics in Physics. M. F. Shlesinger,
 G. M. Zaslavsky \& U. Frish Eds.}, 269, {Springer, Berlin Heidelberg New York, 1995}.

\bibitem{Compte}
   A. Compte, Phys. Rev. E \textbf{53}, 4191 (1996).

\bibitem{Plastino}
   A. R. Plastino  \& A. Plastino,
   Physica A \textbf{222}, 347 (1995).

\bibitem{Jou}
   A. Compte \& D. Jou, J. Phys. A \textbf{29}, 4321 (1996).

\bibitem{Borland}
   L. Borland, Phys. Rev. E \textbf{57}, 6643 (1998).

\bibitem{Upadhyaya}
   A. Upadhyaya, J-P. Rieu, J. Glazier \& Y. Sawada, Physica A
\textbf{293}, 549 (2001).

\bibitem{Bodenschatz}
   K. E. Daniels, C. Beck \& E. Bodenschatz, Physica A \textbf{193}, 208 (2004).

\bibitem{Renzoni}
   P. Douglas, S. Bergamini \& F. Renzoni,  Phys. Rev. Lett. \textbf{96}, 110601 (2006).

\bibitem{Sattin}
   F. Sattin, J. Phys. A: Math. Gen \textbf{36}, 1583 (2003).


\end{thebibliography}
\end{document}